\documentclass[twocolumn,aps,prc,showpacs,superscriptaddress,preprintnumbers,amssymb,floatfix,nofootinbib]{revtex4-2}
\usepackage{newtxtext,newtxmath,booktabs,siunitx}
\usepackage{epsfig,graphics}
\usepackage{graphicx}
\usepackage{dcolumn}
\usepackage{bm}
\usepackage{amsmath}
\usepackage{ulem} 
\usepackage[colorlinks,linkcolor=blue,urlcolor=blue,citecolor=blue]{hyperref}
\usepackage{tikz}
\usepackage{gensymb}
\usepackage{booktabs}
\usepackage{subfigure}
\usetikzlibrary {shapes.geometric}
\usetikzlibrary{shapes,arrows}

\newcommand{\U}{$^{238}$U}
\newcommand{\Au}{$^{197}$Au}

\newcommand{\AuAu}{$^{197}$Au+$^{197}$Au}
\newcommand{\UU}{$^{238}$U+$^{238}$U}

\newcommand{\vtwosq}{v_{2}\{2\}^{2}}
\newcommand{\vthreesq}{v_{3}\{2\}^{2}}
\newcommand{\etwosq}{\varepsilon_{2}^{2}}

\newcommand{\vtwopt}{\langle v_{2}^{2}\delta p_{\mathrm T} \rangle}
\newcommand{\deltapT}{\langle(\delta p_{\mathrm T})^{2}\rangle}
\newcommand{\iebe}{\texttt{iEBE-VISHNU}}
\newcommand{\trento}{\texttt{T\raisebox{-0.5ex}{R}ENTo}}
\newcommand{\Audef}{${\rm Au}_{\rm default}^{\rm WS}$}
\newcommand{\Aunew}{${\rm Au}_{\rm new}^{\rm WS}$}
\newcommand{\Uoct}{U$_{\rm octupole}^{\rm DFT}$}
\newcommand{\Upro}{U$_{\rm prolate}^{\rm DFT}$}
\begin{document}

\title{Testing microscopic nuclear-density inputs for deformation-sensitive observables in relativistic \UU\ and \AuAu\ collisions}

\author{Yuan Li}
\affiliation{Key Laboratory of Nuclear Physics and Ion-beam Application (MOE), Institute of Modern Physics, Fudan University, Shanghai 200433, China}
\affiliation{School of Science, Huzhou Normal University, Huzhou, Zhejiang 313000, China}
\affiliation{Shanghai Research Center for Theoretical Nuclear Physics, NSFC and Fudan University, Shanghai 200438, China.}

\author{Hao-jie Xu}
\email{haojiexu@zjhu.edu.cn}
\affiliation{School of Science, Huzhou Normal University, Huzhou, Zhejiang 313000, China}
\affiliation{Shanghai Research Center for Theoretical Nuclear Physics, NSFC and Fudan University, Shanghai 200438, China.}
\affiliation{Strong-Coupling Physics International Research Laboratory (SPiRL), Huzhou Normal University, Huzhou, Zhejiang 313000, China.}

\author{Dandan Zhang}
\email{ddzhang@itp.ac.cn}
\affiliation{Institute of Theoretical Physics, Chinese Academy of Sciences, Beijing 100871, China}

\author{Guo-Liang Ma}
\email{glma@fudan.edu.cn}
\affiliation{Key Laboratory of Nuclear Physics and Ion-beam Application (MOE), Institute of Modern Physics, Fudan University, Shanghai 200433, China}
\affiliation{Shanghai Research Center for Theoretical Nuclear Physics, NSFC and Fudan University, Shanghai 200438, China.}

\begin{abstract}
Recent STAR measurements of \UU/\AuAu\ ratios provide a sensitive test of nuclear deformation in relativistic heavy-ion collisions, but the elliptic-flow, triangular-flow, and transverse-momentum-related observables are not simultaneously described within existing calculations.  We use three-dimensional lattice covariant density functional theory (CDFT) with pairing correlations to provide microscopic density distributions of both \U\ and \Au\ for event-by-event hydrodynamic simulations.  The CDFT densities substantially improve the description of the ultracentral elliptic-flow ratio, while the octupole-deformed uranium configuration yields a larger triangular-flow ratio than the reflection-symmetric configuration.  In contrast, the transverse-momentum-related ratios remain difficult to describe within the present framework.  We further show that variations of the density of the odd-mass \Au\ reference nucleus can produce changes in the triangular-flow ratio comparable to the separation between the two uranium configurations.  These results demonstrate that both the uranium structure and the reference-nucleus density must be controlled before quantitative nuclear-structure constraints can be extracted from high-energy collisions.
\end{abstract}

\maketitle

\section{Introduction}
The possibility of imaging nuclear deformation with relativistic heavy-ion collisions has recently attracted considerable attention~\cite{Schenke:2020mbo,Giacalone:2021udy,Jia:2021tzt,Magdy:2022cvt,Ryssens:2023fkv,Xu:2024bdh,Wang:2024vjf, Fortier:2023xxy, Giacalone:2024bud,STAR:2024wgy,STAR:2025elk}.  In these collisions, pressure gradients in the quark-gluon plasma (QGP) convert the spatial anisotropies of the initial geometry, characterized by eccentricities $\varepsilon_n$, into final-state anisotropic flow coefficients $v_n$~\cite{Romatschke:2007mq,Song:2010mg,Xu:2016hmp,Zhao:2017yhj}.  For central collisions of deformed nuclei, the intrinsic nuclear shape can therefore leave measurable imprints on observables such as $v_2$, $v_3$, the event-by-event mean transverse momentum $[p_{\mathrm T}]$, and correlations between flow and $[p_{\mathrm T}]$~\cite{Giacalone:2020awm,Jia:2021wbq,Jia:2021qyu}.  Since the collision time is only a few yoctoseconds, high-energy collisions provide a short-time snapshot of the nuclear matter distribution that is complementary to low-energy spectroscopic probes~\cite{STAR:2024wgy}.

The \U\ nucleus is a particularly important system for this purpose.  Its large quadrupole deformation is well established from low-energy nuclear-structure information, including the measured ground-state electric transition rate $B(E2)=12.09\pm0.20~\rm e^2 b^2$~\cite{Raman:2001nnq}.  Higher-order components are less certain.  Recent heavy-ion studies have shown that a finite hexadecapole component of \U\ can modify deformation-sensitive observables~\cite{Ryssens:2023fkv,Xu:2024bdh}, while octupole collectivity is a well-known degree of freedom in reflection-asymmetric nuclei~\cite{Butler:1996zz}.  The covariant density functional theory (CDFT) calculations used below further indicate an octupole-deformed configuration of \U, providing a microscopic input for testing whether the triangular-flow ratio retains sensitivity to nuclear structure beyond the dominant quadrupole deformation~\cite{Zhang:2023slb}.

Recent STAR measurements of deformation-sensitive \UU/\AuAu\ ratios make such a test especially timely~\cite{STAR:2024wgy,STAR:2025elk}.  The elliptic-flow, triangular-flow, and transverse-momentum-related ratios probe different aspects of the initial geometry and its dynamical response.  The central question is whether one consistent set of microscopic density inputs for both \U\ and \Au\ can describe their observed pattern within the same dynamical framework.  A failure of the tested density variations to describe particular observables would then identify where uncertainties beyond the static nuclear-density input remain important.

Answering this question requires a realistic description of the nuclear density distributions used in the heavy-ion initial condition.  Simple deformed Woods--Saxon parameterizations have been widely used, but their surface deformation parameters are not identical to the intrinsic multipole moments obtained from microscopic nuclear-structure calculations~\cite{Hagino:2006fj}.  In particular, the finite hexadecapole component of \U\ can modify the effective quadrupole geometry relevant for heavy-ion observables~\cite{Ryssens:2023fkv,Xu:2024bdh}.  Such differences have already led to ambiguities in the interpretation of the elliptic flow ratio in \UU/\AuAu\ collisions~\cite{STAR:2015mki,Ryssens:2023fkv,Xu:2024bdh}.  Microscopic density profiles from density functional theory are therefore needed to reduce this source of model dependence.

The reference nucleus introduces a further complication.  All \UU/\AuAu\ ratios depend not only on the structure of \U\ but also on the density distribution of \Au.  Unlike the even-even uranium nucleus, \Au\ is an odd-mass nucleus with a triaxial ground-state structure, and its density distribution is less tightly constrained.  Consequently, a change in a \UU/\AuAu\ ratio may reflect the uranium deformation, the Au reference density, or the dynamical response of the collision system.  Quantifying the influence of the Au density is therefore a necessary step toward interpreting these ratios as constraints on the multipole deformation of \U.

In this work, we use event-by-event hydrodynamic simulations to determine which features of the STAR deformation-sensitive ratios can be explained by microscopic nuclear-density inputs.  We employ CDFT densities for both \U\ and \Au\ as the primary inputs.  For \U, we compare an octupole-deformed configuration with $\beta_{20}=0.29$ and $\beta_{30}=0.09$ with a reflection-symmetric prolate configuration that has the same quadrupole deformation.  For \Au, we first use the CDFT triaxial density and then replace it with two Woods--Saxon parameterizations to assess the sensitivity to the reference density.  This comparison allows us to examine separately the effects associated with the uranium configuration and the \Au\ reference density, and to identify observables whose remaining discrepancies motivate further investigation of initial-state fluctuations, size--eccentricity correlations, or hydrodynamic response.

\section{Model and setup}
\label{sec:model}
In this study, \UU\ and \AuAu\ collisions are simulated with the event-by-event (2+1)-dimensional viscous hydrodynamic model \iebe~\cite{Heinz:2005bw, Song:2007ux, Shen:2014vra, Bernhard:2016tnd}, which describes the dynamical evolution of the QGP medium and is coupled to the hadron cascade model {\tt UrQMD} for the subsequent hadronic stage~\cite{Bass:1998ca, Bleicher:1999xi}.  The initial conditions are generated with the \texttt{$\text{T}_{\text{R}}\text{ENTo}$} model~\cite{Bernhard:2016tnd, Moreland:2014oya} for a given nuclear density distribution.  All parameters in the \iebe\ simulations are taken from Ref.~\cite{Bernhard:2019bmu}, except for the overall normalization, which is adjusted to reproduce the charged-particle multiplicity.

The nuclear densities are calculated with the point-coupling relativistic energy density functional PC-PK1.  The density profiles for \U\ are obtained from CDFT calculations in a three-dimensional lattice space with $L_x\times L_y\times L_z=24\times24\times30$ fm$^{3}$.  The deformation parameters at the equilibrium minimum are $\beta_{20}=0.29$ and $\beta_{30}=0.09$~\footnote{These deformation parameters differ from those used in conventional Woods--Saxon parameterizations. The latter are denoted by \(\beta^{\rm WS}\).}, which we refer to as the octupole configuration.  To assess the change associated with the octupole-deformed solution, we compare this density with the minimum along the $\beta_{30}=0$ direction, referred to as the reflection-symmetric prolate configuration.  Further details of the CDFT calculations for \U\ can be found in Ref.~\cite{Zhang:2023slb}. 

\begin{figure}[htbp]
\centering
\includegraphics[width=0.45\textwidth]{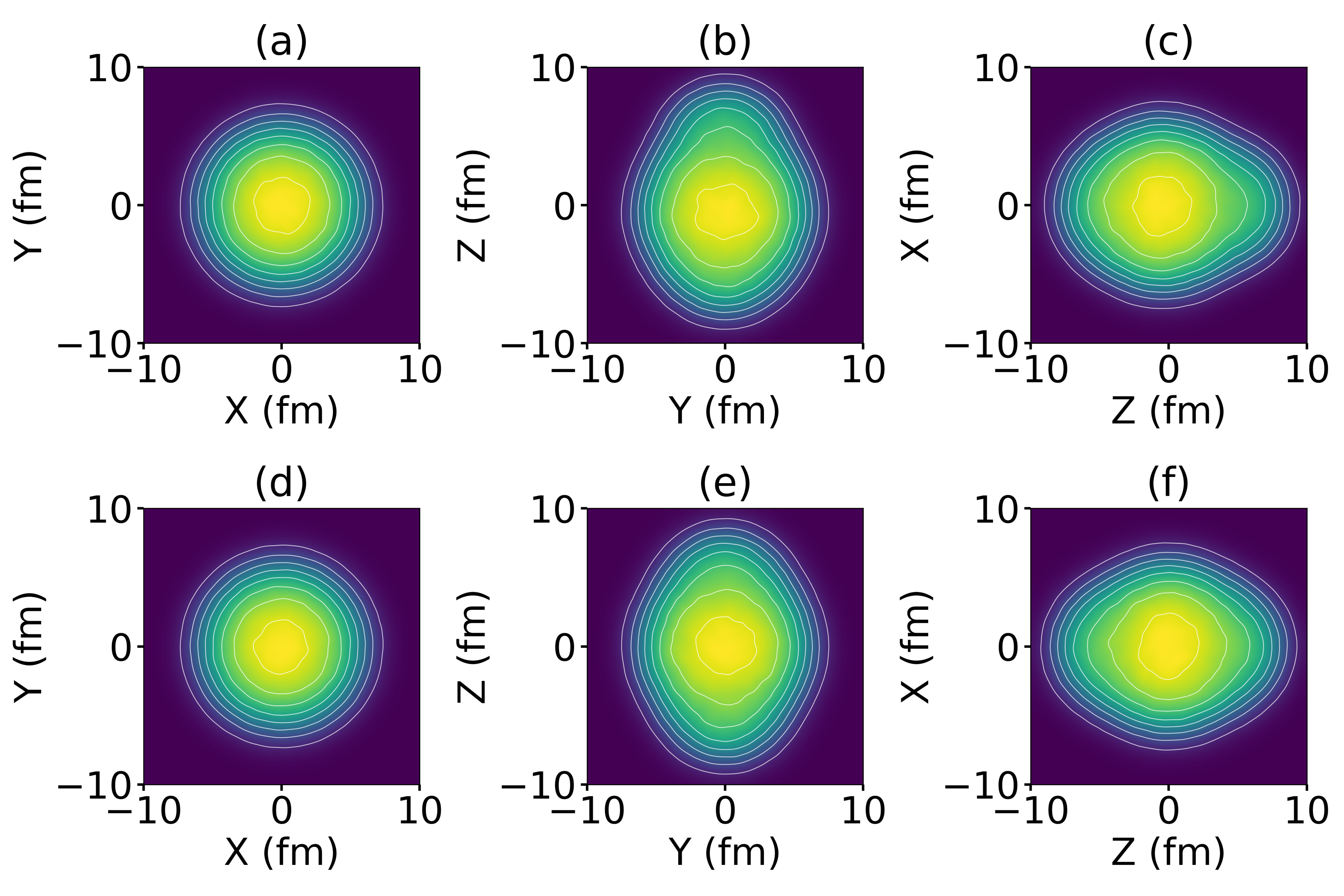}
\caption{(Color online) Density distributions of \U\ from CDFT calculations.  The upper panels show the octupole-deformed configuration with $\beta_{20}=0.29$ and $\beta_{30}=0.09$, while the lower panels show the reflection-symmetric prolate configuration with $\beta_{20}=0.29$ and $\beta_{30}=0$.}
\label{densityprofiles}
\end{figure}

The density profile of \Au\ is calculated within the same CDFT framework.  The resulting ground-state density has a triaxial quadrupole deformation characterized by $\beta_{2}=0.13$ and $\gamma=44\degree$.  The CDFT density distributions of both \U\ and \Au\ are therefore used as the primary nuclear-structure inputs in the calculations presented below. 
The resulting \U\ density profiles exhibit pronounced differences between the octupole-deformed and reflection-symmetric prolate configurations, as illustrated in Fig.~\ref{densityprofiles}.

For each \UU\ and \AuAu\ configuration, approximately $10^{6}$ hydrodynamic events are generated, with each event followed by 10 oversamplings using the \texttt{UrQMD} afterburner.  The flow observables are calculated using both the standard Q-cumulant method~\cite{Bilandzic:2010jr, Bilandzic:2013kga} and a pseudorapidity-separated subevent method, which give consistent results.  The standard Q-cumulant method is applied to charged particles with transverse momentum $0.2 < p_{T} < 2~\mathrm{GeV}/c$ and pseudorapidity $|\eta| < 2$.

We next define the observables used in the analysis.  The flow coefficients and flow--$p_{\mathrm T}$ correlations are calculated with the two-particle cumulant method,
\begin{subequations}\label{eq:cn}
    \begin{align}
        v_n\{2\} &\equiv \sqrt{c_n\{2\}} \equiv \sqrt{\langle \langle 2 \rangle_{n,-n}\rangle}\,, \\
        \langle v_{n}^{2} \delta p_{\mathrm{T}} \rangle &\equiv \operatorname{Cov}\left(v_n^2,\left[p_{\mathrm T}\right]\right) \nonumber \\
        &= \langle v_{n}^{2} [p_{\mathrm T}] \rangle - \langle v_{n}^{2} \rangle \langle [p_{\mathrm T}] \rangle\,,
    \end{align}
\end{subequations}
where \(\langle \langle m \rangle_{n_1,n_2,\ldots,n_m} \rangle\) denotes the correlator
\begin{equation}
\langle \langle m \rangle_{n_1,n_2,\ldots,n_m} \rangle \equiv \left\langle e^{i(n_1\varphi_1+n_2\varphi_2+\cdots+n_m\varphi_m)} \right\rangle.
\end{equation}
The inner $\langle\cdots\rangle$ indicates an average over particles of interest within a single event, while the outer average is taken over events~\cite{Bilandzic:2010jr, Bilandzic:2013kga}.  The event-wise mean transverse momentum is denoted by $[p_{\mathrm T}]$, and $\operatorname{Cov}\left(v_n^2,[p_{\mathrm T}]\right)$ is the covariance between $v_n^2$ and $[p_{\mathrm T}]$.

In addition to $\langle v_n^2\delta p_{\mathrm T}\rangle$, the variance of $[p_{\mathrm T}]$, $\deltapT\equiv\sigma_{p_{\mathrm T}}^{2}$, is also sensitive to nuclear deformation effects~\cite{Giacalone:2020awm, Jia:2021wbq, Jia:2021qyu}. Here, $\delta p_{\mathrm T} = [p_{\mathrm T}] - \left\langle[p_{\mathrm T}]\right\rangle_{\text{event}}$ is the fluctuation of $[p_{\mathrm T}]$ in a given centrality range, and $\langle\cdots\rangle$ denotes the average over an ensemble of events.

The mapping from nuclear deformation to final-state observables is influenced by the medium evolution in relativistic heavy-ion collisions, which introduces theoretical uncertainties.  These uncertainties can be reduced by comparing similar collision systems, for which many systematic effects largely cancel.  A prominent example is the comparison of isobaric Ru+Ru and Zr+Zr collisions~\cite{Li:2019kkh, STAR:2021mii, Koch:2016pzl}.  In this work, we use \AuAu\ collisions as the reference for \UU\ collisions and construct the relative observables
\begin{equation}
R(X)=\frac{X_{\mathrm{UU}}}{X_{\mathrm{AuAu}}},
\end{equation}
where $X$ stands for a given observable, e.g., $v_n^{2}\{2\}$, $\vtwopt$, or $\deltapT$.

Our primary calculations employ CDFT density profiles for both \U\ and \Au\ to avoid the model dependence associated with inconsistent structural descriptions.  To quantify the sensitivity to the \Au\ reference density, we perform additional calculations using the two Woods--Saxon parameterizations listed in Table~\ref{tab:deformation}.  These calculations test whether variations of the Au density can account for the remaining differences between the CDFT predictions and the experimental data.

\begin{table}[htb]
    \caption{Woods--Saxon parameters employed for \Au\ in the sensitivity study.  The parameter sets \Audef\ and \Aunew\ correspond to the two Au density profiles discussed in the text.}
    \label{tab:deformation}

    \renewcommand{\arraystretch}{1.4}
    
    \begin{ruledtabular}
    \begin{tabular}{lcccc}
        Configurations & $R_{0}$ (fm) & $a$ (fm) & $\beta_{2}^{\rm WS}$ & $\beta_{4}^{\rm WS}$ \\
        \hline
        \Audef~\cite{STAR:2024wgy} & 6.38 & 0.535 & -0.14 & 0.00 \\
        \Aunew~\cite{Shou:2014eya} & 6.42 & 0.410 & -0.14 & 0.00 \\
    \end{tabular}
    \end{ruledtabular}
\end{table}

\section{Results and discussion}
\label{sec:results}
\begin{figure*}[htbp]
\centering
\includegraphics[width=0.48\textwidth]{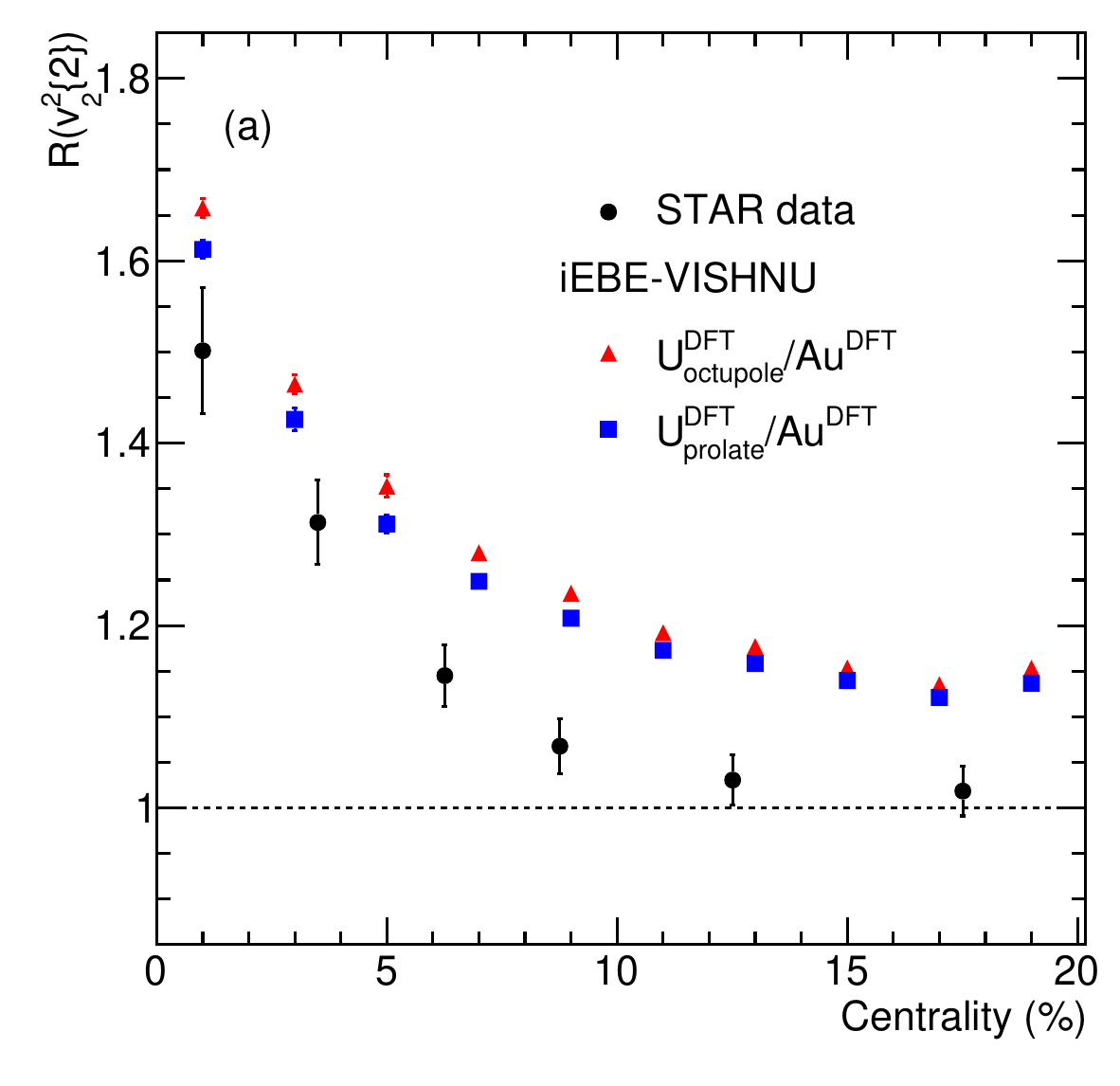}
\includegraphics[width=0.48\textwidth]{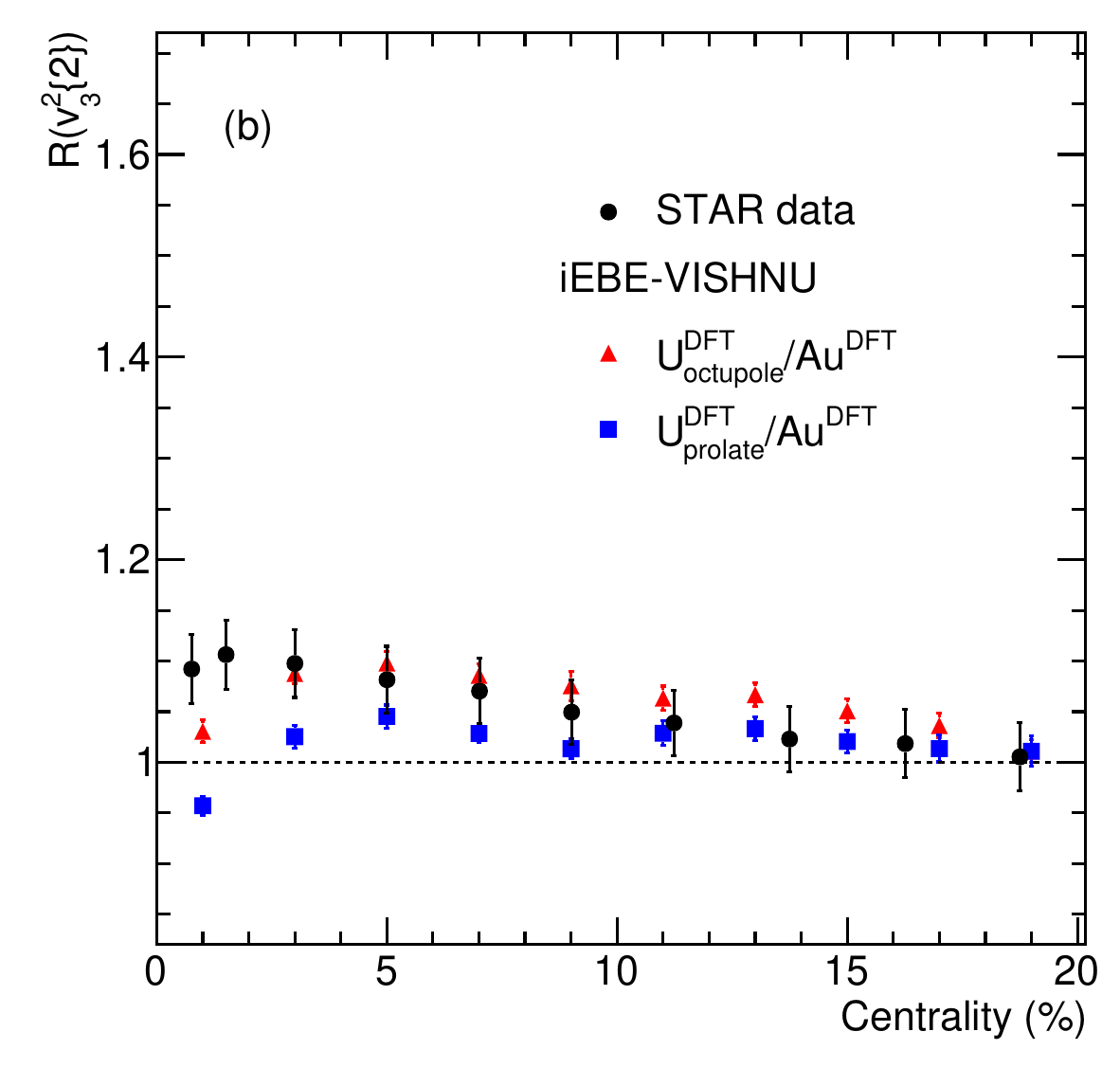}
\includegraphics[width=0.48\textwidth]{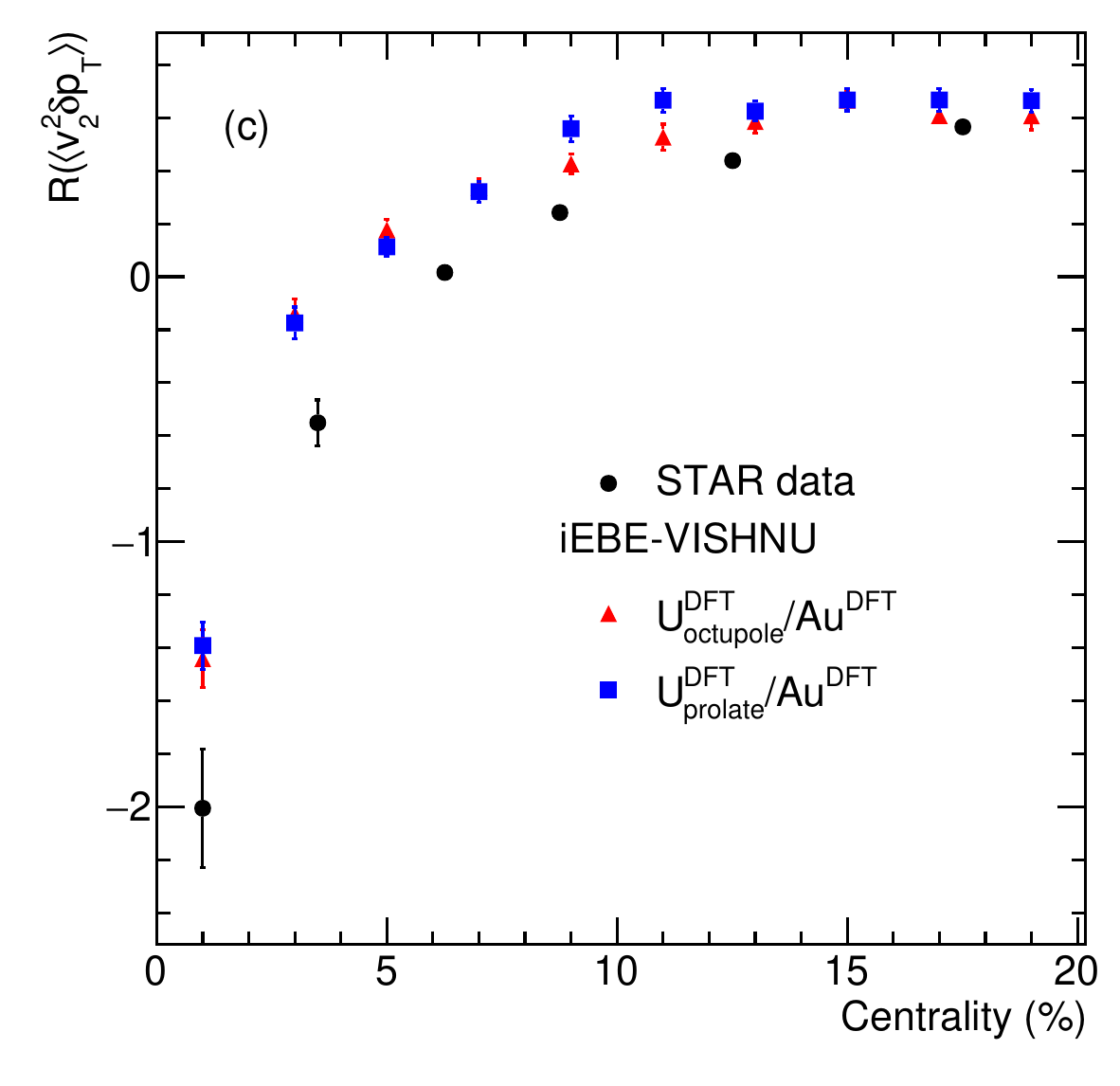}
\includegraphics[width=0.48\textwidth]{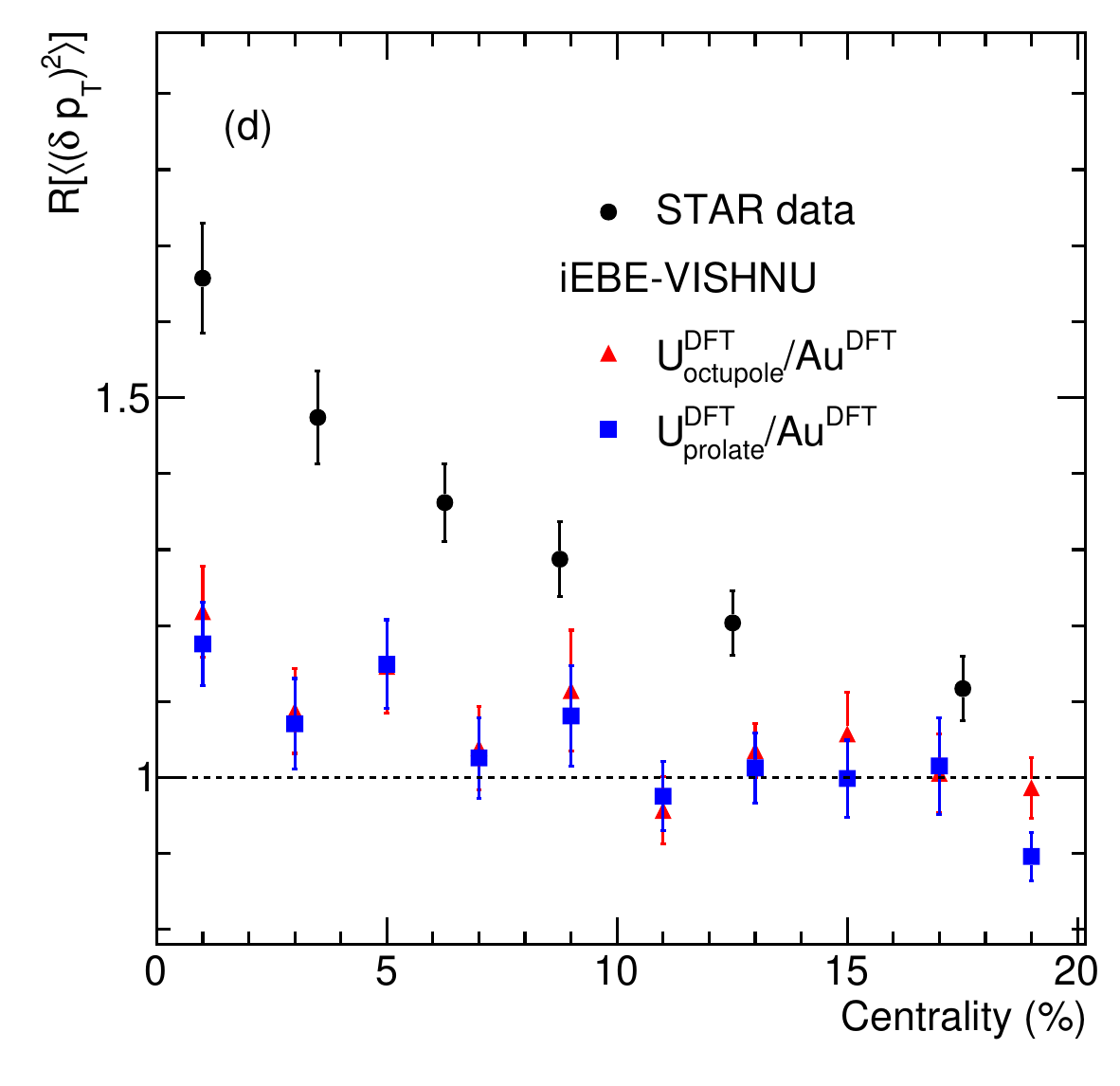}
\caption{(Color online) Centrality dependence of the \UU/\AuAu\ ratios obtained from \iebe\ simulations using CDFT density distributions for both nuclei.  Panels show (a) $R(v_{2}^{2}\{2\})$, (b) $R(v_{3}^{2}\{2\})$, (c) $R(\langle v_{2}^{2}\delta p_{\mathrm T}\rangle)$, and (d) $R(\langle(\delta p_{\mathrm T})^{2}\rangle)$.  Red triangles and blue squares denote the octupole and reflection-symmetric prolate configurations of \U, respectively.  Experimental data from STAR are taken from Refs.~\cite{STAR:2024wgy,STAR:2025elk}.}
\label{fig:dft_ratios}
\end{figure*}

We now compare the deformation-sensitive \UU/\AuAu\ ratios obtained with different nuclear-density inputs.  The goal is to determine which features of the STAR measurements are explained once both nuclei are described within the same microscopic CDFT framework and which features remain unresolved.  The first comparison, shown in Fig.~\ref{fig:dft_ratios}, uses CDFT density profiles for both \U\ and \Au.  This provides the most direct test of the role of improved input densities in the observed ratios.

The four observables shown in Fig.~\ref{fig:dft_ratios} probe different aspects of the initial geometry.  The elliptic flow ratio $R(v_2^2\{2\})$ is primarily sensitive to the quadrupole geometry, whereas the triangular flow ratio $R(v_3^2\{2\})$ provides complementary sensitivity to octupole-induced triangularity.  The flow--mean-transverse-momentum correlation $R(\langle v_2^2\delta p_{\mathrm T}\rangle)$ and the transverse-momentum fluctuation ratio $R(\langle(\delta p_{\mathrm T})^2\rangle)$ are more differential tests: they depend not only on eccentricity fluctuations but also on transverse-size fluctuations and their correlations.  A simultaneous description of all four ratios therefore tests both the nuclear density input and the dynamical mapping from the initial state to the final observables.

\begin{figure*}[htbp]
\centering
\includegraphics[width=0.48\textwidth]{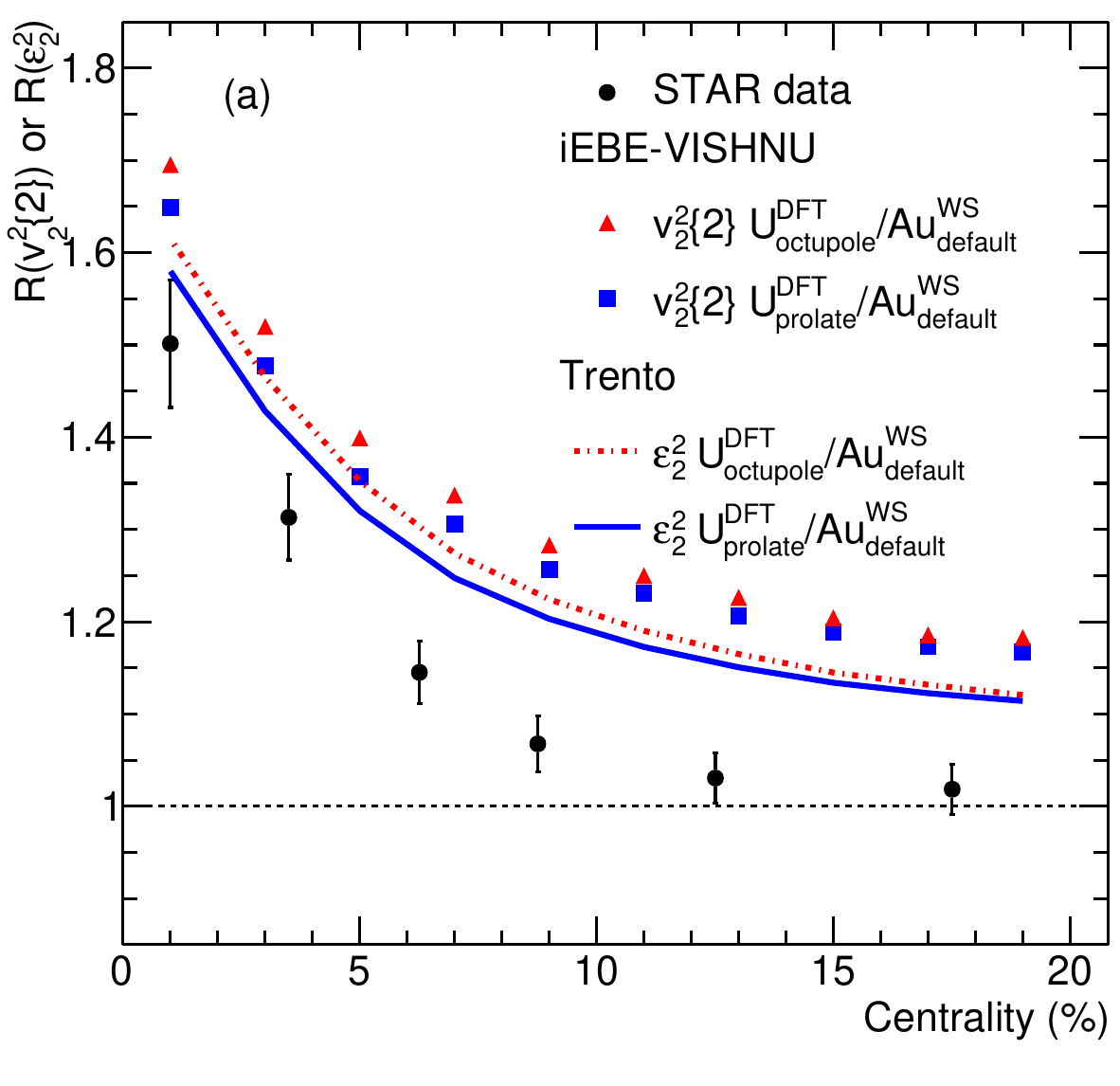}
\includegraphics[width=0.48\textwidth]{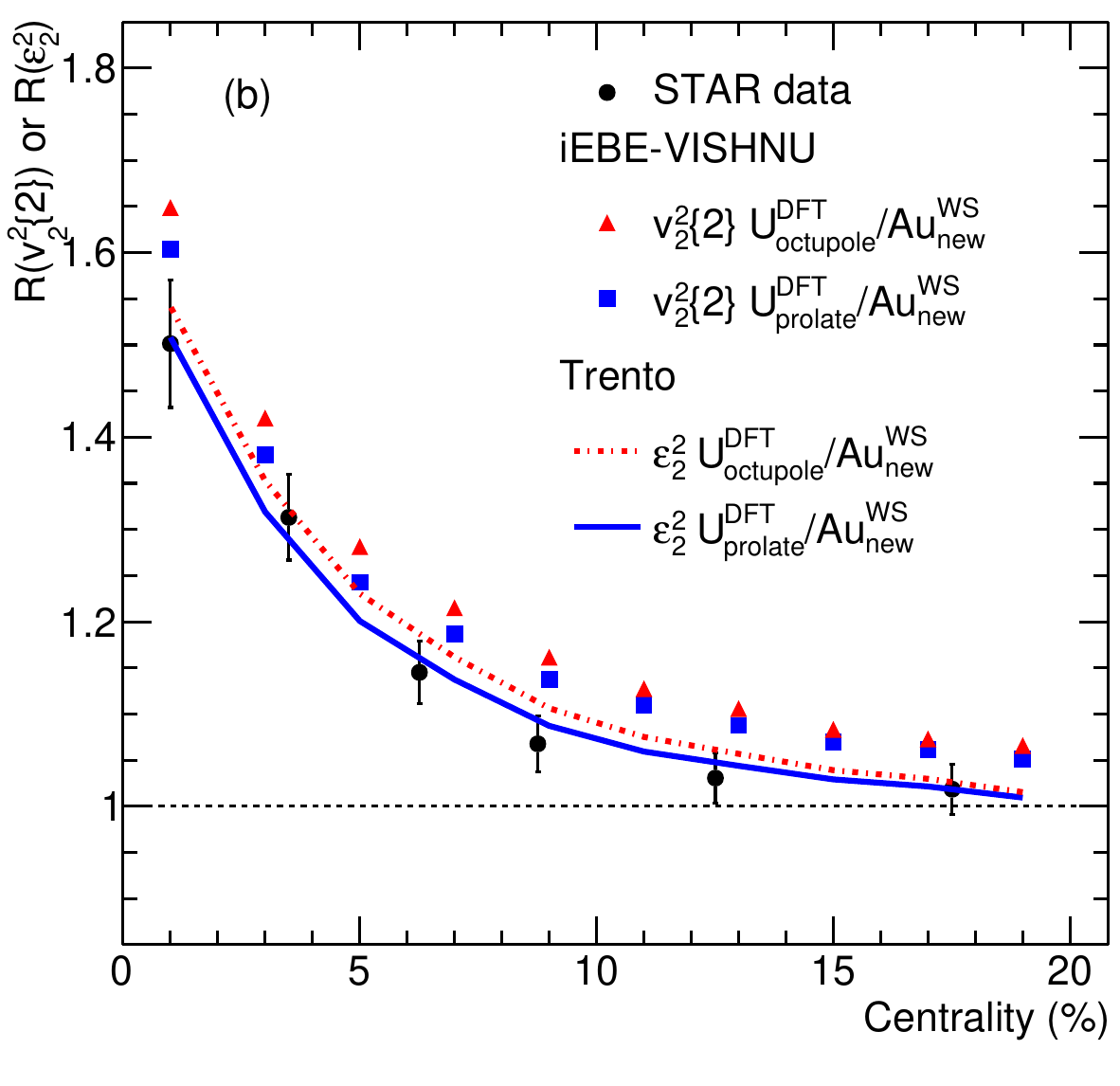}
\caption{(Color online) Centrality dependence of $R(v_{2}^{2}\{2\})$ and the corresponding initial-state ratio $R(\varepsilon_{2}^{2})$ between U+U and Au+Au collisions.  Panels (a) and (b) use the \Audef\ and \Aunew\ Woods--Saxon parameter sets for \Au, respectively.  Final-state results are obtained from \iebe\ simulations, while the initial-state ratios are obtained from \trento.  Experimental data from STAR are taken from Ref.~\cite{STAR:2024wgy}.}
\label{v2ratio}
\end{figure*}

\begin{figure*}[htbp]
\centering
\includegraphics[width=0.48\textwidth]{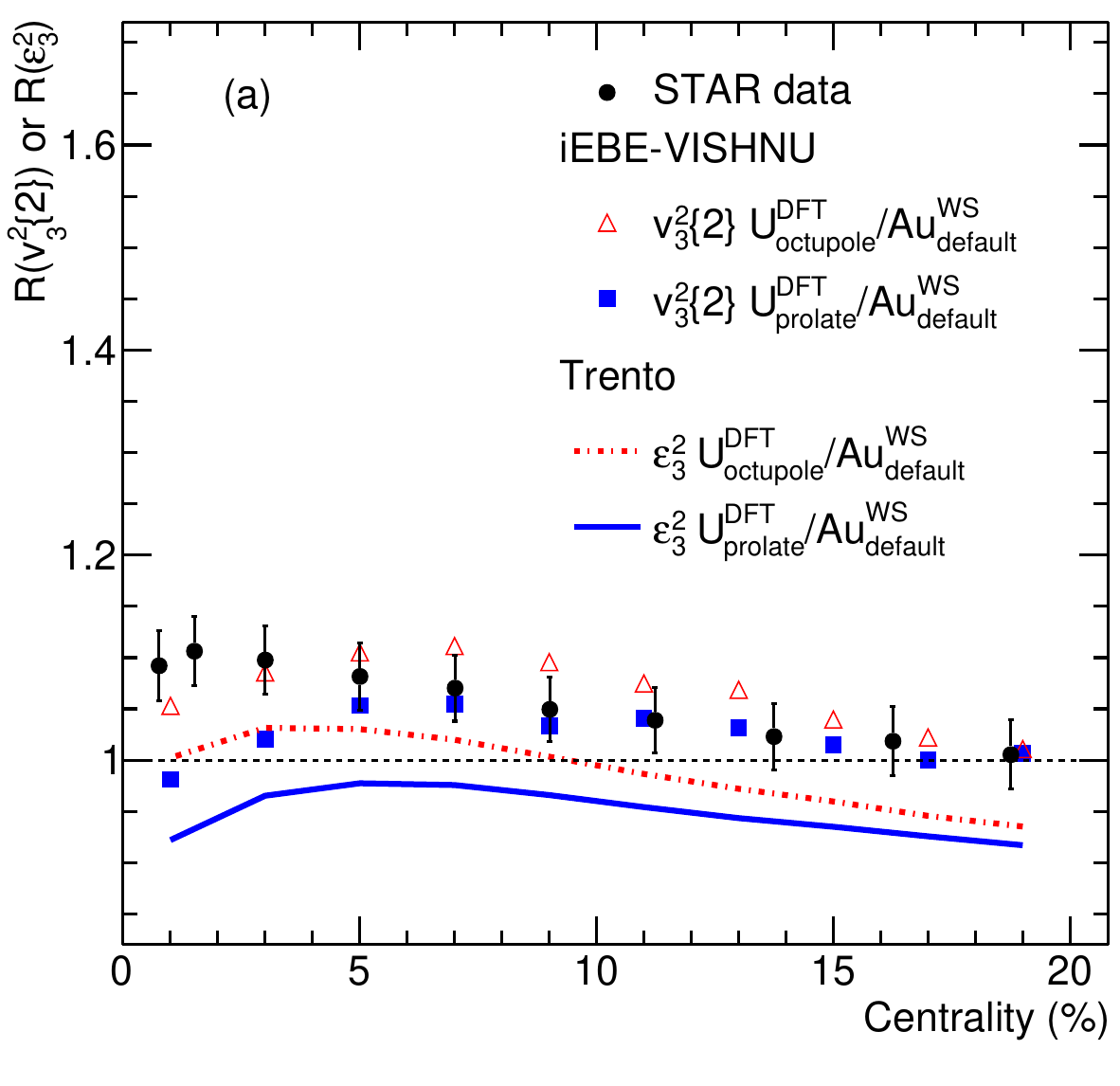}
\includegraphics[width=0.48\textwidth]{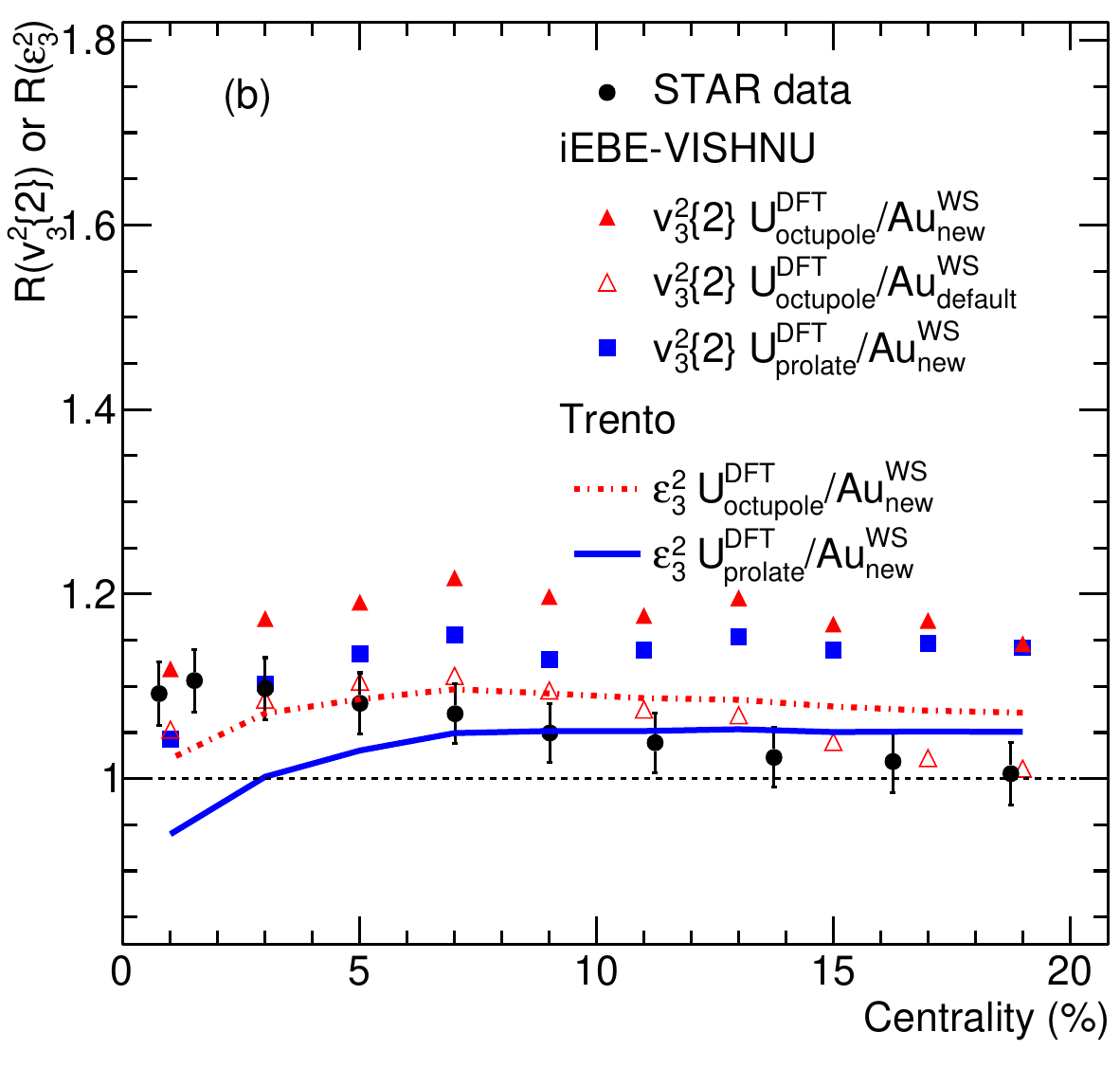}
\caption{(Color online) Centrality dependence of $R(v_{3}^{2}\{2\})$ and the corresponding initial-state ratio $R(\varepsilon_{3}^{2})$ between U+U and Au+Au collisions.  Panels (a) and (b) use the \Audef\ and \Aunew\ Woods--Saxon parameter sets for \Au, respectively.  Final-state results are obtained from \iebe\ simulations, while the initial-state ratios are obtained from \trento.  Experimental data from STAR are taken from Ref.~\cite{STAR:2025elk}.}
\label{v3ratio}
\end{figure*}

\begin{figure}[htbp]
\centering
\includegraphics[width=0.45\textwidth]{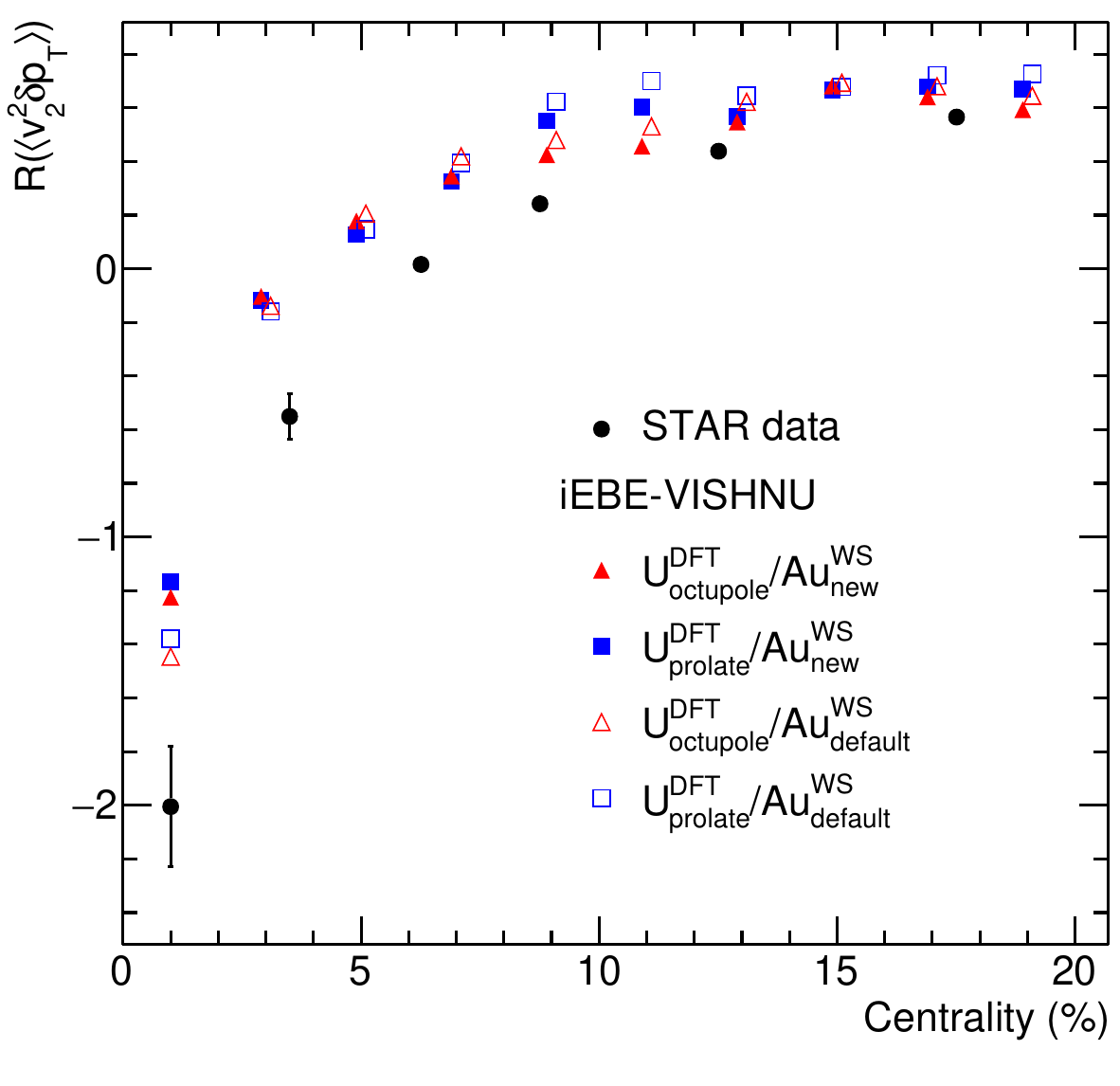}
\caption{(Color online) Centrality dependence of the $R(\vtwopt)$ ratio between U+U and Au+Au collisions obtained from \iebe\ simulations.  Experimental data from STAR are taken from Ref.~\cite{STAR:2024wgy}.}
\label{v2dptratio}
\end{figure}

\begin{figure}[htbp]
\centering
\includegraphics[width=0.45\textwidth]{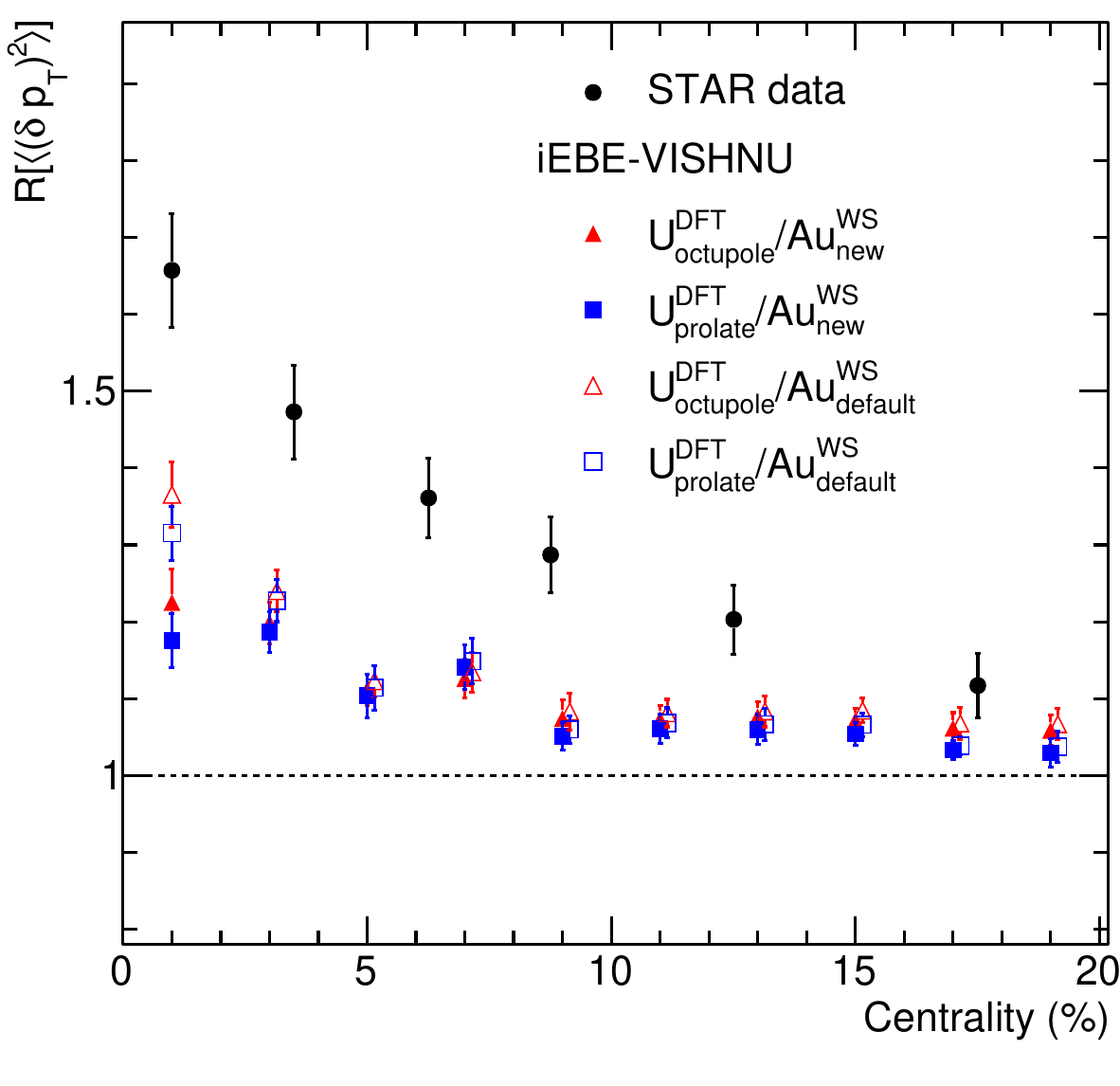}
\caption{(Color online) Centrality dependence of the $R[\deltapT]$ ratio between U+U and Au+Au collisions obtained from \iebe\ simulations.  Experimental data from STAR are taken from Ref.~\cite{STAR:2024wgy}.}
\label{dptratio}
\end{figure}

Figure~\ref{fig:dft_ratios}(a) shows $R(\vtwosq)$ for the octupole and prolate configurations of \U.  Since the two uranium configurations have the same quadrupole deformation, $\beta_{20}=0.29$, they give similar values of $R(\vtwosq)$.  The octupole configuration produces a slightly larger ratio than the prolate configuration, but the difference remains small throughout the centrality range considered.

Although the intrinsic CDFT quadrupole deformation $\beta_{20}=0.29$ is numerically identical to the Woods--Saxon parameter $\beta_{2}^{\rm WS}=0.29$ often used in previous studies, the two values do not correspond to equivalent density distributions.  In the CDFT density, the finite hexadecapole deformation, $\beta_{40}=0.16$, modifies the surface geometry such that the corresponding effective Woods--Saxon quadrupole deformation is substantially smaller than the intrinsic $\beta_{20}$~\cite{Ryssens:2023fkv,Xu:2024bdh}.  As a result, the pronounced overestimation of $R(\vtwosq)$ in ultracentral collisions, commonly referred to as the ultracentral $v_2$ puzzle, is largely resolved in the present calculations.

The corresponding triangular flow ratio is shown in Fig.~\ref{fig:dft_ratios}(b).  In contrast to $R(\vtwosq)$, a clear separation is observed between the octupole and prolate uranium configurations.  The finite $\beta_{30}$ of the octupole configuration enhances $R(\vthreesq)$ by approximately $0.2$ relative to the reflection-symmetric prolate configuration.  In the most central collisions, the octupole configuration gives a better description of the magnitude of the experimental ratio, whereas the prolate configuration predicts a smaller value.

This difference demonstrates that $R(\vthreesq)$ retains sensitivity to the octupole component of the uranium density.  However, the separation between the two configurations is comparable to the present experimental uncertainty and, as discussed below, is also affected by uncertainties in the density distribution of the Au reference nucleus.  The triangular flow ratio alone therefore does not yet provide an unambiguous quantitative constraint on the octupole deformation of \U.

Figure~\ref{fig:dft_ratios}(c) presents $R(\vtwopt)$.  The calculation qualitatively reproduces the increase of the ratio from central to more peripheral collisions.  Quantitatively, however, the calculated values are systematically larger than the experimental measurements.  A similar tension is observed for the mean-transverse-momentum fluctuation ratio shown in Fig.~\ref{fig:dft_ratios}(d).  The experimental $R[\deltapT]$ exhibits a pronounced enhancement toward the most central collisions, whereas the CDFT-based calculations predict a considerably weaker enhancement and remain systematically below the data in the central region.

Taken together, Fig.~\ref{fig:dft_ratios} shows that the use of consistent microscopic densities substantially improves the description of the flow ratios but does not yield a simultaneous description of all four observables within the present framework.  The CDFT inputs reduce a major part of the previously observed ultracentral $v_2$ tension, and the two uranium configurations remain distinguishable in $R(\vthreesq)$. 
At the same time, the transverse-momentum-related ratios are not described.  The remaining discrepancies are therefore not removed by the uranium density variations considered here.

A natural next question is whether the remaining differences are amplified or masked by the Au reference system.  Although the CDFT calculation provides a microscopic Au density, \Au\ is an odd-mass triaxial nucleus, and its density is less constrained than that of an even-even reference nucleus.  Since every measured ratio is normalized by the \AuAu\ result, changes in the Au surface profile can propagate into the inferred uranium deformation signal.  To quantify this effect, we keep the CDFT densities of \U\ fixed and replace the CDFT density of \Au\ with the two Woods--Saxon parameterizations listed in Table~\ref{tab:deformation}.

Figures~\ref{v2ratio}(a) and (b) show $R(\vtwosq)$ calculated with the \Audef\ and \Aunew\ parameter sets, respectively.  Symbols and curves denote the final-state flow ratio $R(\vtwosq)$ and the corresponding initial-state eccentricity ratio $R(\etwosq)$.  The close agreement between the \trento\ and \iebe\ ratios in the 0--20\% centrality range is consistent with an approximate cancellation of the hydrodynamic response coefficients between the two collision systems~\cite{Qiu:2011iv}.  For a given Au parameterization, the octupole and prolate uranium configurations give nearly identical values of $R(\vtwosq)$, as expected from their common quadrupole deformation $\beta_{20}=0.29$.  However, the absolute magnitude of the ratio depends strongly on the Au density.  The commonly employed \Audef\ parameter set produces a systematic overestimation of $R(\vtwosq)$ by approximately $0.1$, as shown in Fig.~\ref{v2ratio}(a).  The \Aunew\ parameterization, which has a smaller surface diffuseness and was extracted by explicitly accounting for nuclear deformation in the analysis of electron-scattering information~\cite{Shou:2014eya}, reduces this overestimation and gives better agreement with the STAR data, as shown in Fig.~\ref{v2ratio}(b).  This comparison demonstrates that the quantitative value of the elliptic flow ratio is sensitive to the surface profile of the Au reference nucleus.

Figures~\ref{v3ratio}(a) and (b) present the corresponding $R(\vthreesq)$ results obtained with the \Audef\ and \Aunew\ parameter sets.  For either Au density, the octupole-deformed uranium configuration gives a larger triangular-flow ratio than the prolate configuration, showing that the separation between the two uranium configurations persists when the reference density is varied.  The absolute magnitude of $R(\vthreesq)$, however, differs substantially between the two Au parameterizations.  The \Audef\ parameter set describes the STAR measurements for most centrality intervals except the most central 0--2\% region, whereas the \Aunew\ set improves the agreement in the 0--2\% interval but overestimates the data at larger centralities.  Moreover, the result obtained with the \Uoct\ configuration and the \Audef\ density can be similar to that obtained with the \Upro\ configuration and the \Aunew\ density.  The spread induced by these two representative Au density profiles is therefore comparable to the separation between the two uranium configurations, which limits the extraction of $\beta_{30}$ from $R(\vthreesq)$ alone.

Figures~\ref{v2dptratio} and \ref{dptratio} show the corresponding results for $R(\vtwopt)$ and $R[\deltapT]$, respectively.  For a fixed Au density, the two uranium configurations again yield nearly identical results, showing little sensitivity to the difference between the configurations considered here.  Changing the Au Woods--Saxon parameterization increases the magnitude of $R(\vtwopt)$ but decreases $R[\deltapT]$, demonstrating that the two transverse-momentum-related observables respond differently to variations of the reference density.  Neither Woods--Saxon parameterization provides a simultaneous description of the two observables.  The calculated $R(\vtwopt)$ remains systematically larger than the experimental measurement, whereas $R[\deltapT]$ remains below the data in central collisions.  Together with the CDFT-based results in Fig.~\ref{fig:dft_ratios}, this shows that the disagreement in the transverse-momentum sector cannot be removed by varying the Au surface diffuseness alone.

The comparison between the CDFT and Woods--Saxon descriptions of \Au\ demonstrates that the reference-nucleus density has a sizable impact on the flow ratios, particularly $R(\vtwosq)$ and $R(\vthreesq)$.  For $R(\vthreesq)$, the Au-density effect can be comparable to the difference between the octupole and reflection-symmetric uranium configurations.  A quantitative constraint on $\beta_{30}$ from triangular flow therefore requires better control of the Au reference density.  At the same time, variations of the Au density do not resolve the simultaneous disagreement of the transverse-momentum-related observables.  The persistent tension motivates further investigation of uncertainties beyond the static nuclear density, such as the event-by-event fluctuation structure of the initial state, the correlation between eccentricity and transverse size, and their subsequent hydrodynamic response.  In this context, recent measurements of $v_0(p_{\mathrm T})$ may provide complementary information on transverse-momentum fluctuations and correlations~\cite{ATLAS:2025ztg,ALICE:2025iud}.

We note that the present study uses the intrinsic CDFT density, as is commonly done in deformation-based event-generator calculations. Although the exact laboratory-frame density of the $0^+$ ground state is spherical after quantum symmetry restoration~\cite{Dobaczewski:2025rdi}, recent investigations suggest that the effective geometry probed in relativistic collisions can still retain intrinsic deformation~\cite{Ke:2025tyv,Xu:2026llz,Otsuka:2026evp}. For the strongly deformed \U\ nucleus, such restoration effects are expected to be small, and the intrinsic-density treatment should provide a reasonable baseline. A quantitative investigation of quantum-restoration corrections is left for future work.

\section{Summary}
\label{sec:summary}

We have used microscopic CDFT densities in event-by-event hydrodynamic simulations to examine how the STAR \UU/\AuAu\ ratios respond to the initial nuclear geometry.  The purpose is to determine which features of these deformation-sensitive observables can be described by consistent microscopic density inputs and which discrepancies remain within the present framework.

The elliptic flow ratio is mainly controlled by the effective quadrupole geometry.  The two uranium configurations considered here, with and without octupole deformation but with the same $\beta_{20}$, give nearly identical $R(\vtwosq)$.  The CDFT densities also largely reduce the ultracentral $R(\vtwosq)$ tension, showing that the full microscopic density profile, not only a single deformation parameter, is important for interpreting the quadrupole signal.

The triangular flow ratio provides a complementary probe of the octupole component of \U.  The octupole-deformed uranium density gives a visibly larger $R(\vthreesq)$ than the reflection-symmetric density.  At the same time, replacing the CDFT Au density by Woods--Saxon parameterizations changes $R(\vthreesq)$ by a comparable amount, demonstrating that quantitative constraints on $\beta_{30}$ require better control of the \Au\ reference density.

The transverse-momentum-related ratios, $R(\vtwopt)$ and $R[\deltapT]$, are not simultaneously described by the density variations considered here.  Their remaining discrepancies motivate further investigation of size fluctuations, size--eccentricity correlations, and the hydrodynamic response of transverse-momentum fluctuations.

Our results show that the STAR \UU/\AuAu\ ratios provide complementary sensitivity to the effective quadrupole geometry, the octupole-deformed uranium configuration, the \Au\ reference density, and fluctuation-driven dynamics.  The observed degeneracies also demonstrate that these ingredients must be controlled jointly before robust nuclear-structure constraints can be extracted from relativistic heavy-ion collisions.

\section*{Acknowledgments}
This work is supported in part by the National Natural Science Foundation of China under Grants No.~12275082, No.~12547102, No.~12147101, and No.~12325507, the National Key Research and Development Program of China under Grant No.~2022YFA1604900, and the Postdoctoral Fellowship Program and China Postdoctoral Science Foundation under Grant No.~BX20250170.

\bibliography{refv2}

\end{document}